\documentclass[aps,prl,twocolumn,notitlepage,superscriptaddress,showpacs,10pt,longbibliography,floatfix]{revtex4-1} %,linenumbers
\usepackage{amsmath}    % need for subequations
\usepackage{amsfonts}
\usepackage{bm}
\usepackage{amssymb}
\usepackage{graphicx}   % need for figures
\usepackage[usenames,dvipsnames]{xcolor}      % use if color is used in text
\usepackage{subcaption}  % use for side-by-side figures
\usepackage{comment}
\usepackage{siunitx}
\usepackage{enumitem}
\usepackage{upgreek}
\usepackage{mathtools}                                  %I just always load this?
\usepackage{amsthm}                                     %For theorem environments
\usepackage{dsfont}                                     %Fancy 1 for identity matrix
\usepackage{lipsum}                                     %Dummy text until abstract is ready
\usepackage{hyperref}                                   %So things link to each other
\hypersetup{
    colorlinks=true, 
    linkcolor=blue,
    citecolor=blue,
    urlcolor=blue
}
\newtheoremstyle{customtheorem}
        {}{}{}{}                                        %Various spacing options
        {\bfseries}{}{0.5em}                            %Theorem head font and spacing
        {#1 #2:~\thmnote{#3}}                           %Colon instead of period
\theoremstyle{customtheorem}
\usepackage{soul}
\usepackage[normalem]{ulem}

\mathchardef\mhyphen="2D

\newcommand{\one}{\mathds{1}}       %Looks like \mathbb{1} should look
\begin{document}

\title{Band Unfolding via the Quadratic Pseudospectrum}
%\title{Local markers for topological crystalline materials}

\author{Christopher A.\ Bairnsfather}
\affiliation{Department of Mathematics, Purdue University, West Lafayette, Indiana, 47907, USA}

\author{Ralph M.\ Kaufmann}
\affiliation{Department of Mathematics, Purdue University, West Lafayette, Indiana, 47907, USA}
\affiliation{Department of Physics and Astronomy, Purdue University, West Lafayette, Indiana, 47907, USA}

\author{Terry A.\ Loring}
%\email[]{loring@math.unm.edu}
\affiliation{Department of Mathematics and Statistics, University of New Mexico, Albuquerque, New Mexico 87131, USA}

\author{Alexander Cerjan}
\email[]{awcerja@sandia.gov}
\affiliation{Center for Integrated Nanotechnologies, Sandia National Laboratories, Albuquerque, New Mexico 87185, USA}

\date{\today}

\begin{abstract}
    Band theory provides the foundation for understanding electronic structure in crystalline materials, but its reliance on exact translational symmetry limits its applicability to systems with defects, disorder, incommensurate modulations, or large unit cells. Here, we introduce a band unfolding framework that directly generalizes traditional band theory to systems where exact periodicity is absent, and which remains well-defined for both aperiodic and finite systems. To do so, we employ a pseudospectral approach to identify approximate joint eigenvectors of a system's Hamiltonian and translation operators, thereby yielding an unfolded band structure whose features are directly connected to the manifestation of approximate extended states simultaneously localized in energy and crystalline momentum. To reveal bulk-only spectral phenomena in finite systems, we further show that this pseudospectral framework naturally accommodates additional operators that suppress contributions from boundary-localized states, enabling the systematic isolation of intrinsic bulk behavior. We benchmark the scheme on several representative systems in one and two dimensions, including a Fibonacci chain, where our approach is able to both reveal a dispersive envelope while preserving the underlying hierarchy of spectral gaps. Looking forward, this pseudospectral approach may yield a broad framework for predicting momentum-resolved material responses in aperiodic, disordered, and finite systems where conventional band-theoretic methods are not applicable.
\end{abstract}

\maketitle

%%% Introduction %%%
Band theory is one of the foundational tools in the study of natural and artificial materials, from which a wide variety of material properties can be determined, including transport \cite{ashcroftMermin}, optical responses \cite{onida_electronic_2002}, and material topology \cite{RevModPhys.90.015001}. However, this reliance upon band theory also presents a challenge for understanding materials with large unit cells, such as those found in substitutional alloys \cite{turek_electronic_2013} and twisted van der Waals materials \cite{andrei_graphene_2020,he_moire_2021,nuckolls_microscopic_2024}, as well as materials that lack crystalline symmetry altogether, such as quasicrystals \cite{steurer_twenty_2004,tsai_icosahedral_2008,dubois_properties-_2012}. In systems that are periodic on a long length scale, the Brillouin zone becomes correspondingly small and the resulting electronic bands are densely folded, complicating the band structure's interpretation and presenting challenges for predicting material properties. In such systems, an unfolded band-like structure can be constructed by projecting the Bloch eigenstates of the supercell $|\psi_{\mathbf{k}_\textrm{sc}}\rangle$ onto Bloch states associated with a reference primitive cell $|\psi_{\mathbf{k}_\textrm{pc}}\rangle$, yielding spectral weights $|\langle \psi_{\mathbf{k}_\textrm{sc}} | \psi_{\mathbf{k}_\textrm{pc}}\rangle|^2$ that reveal the system's momentum-resolved character \cite{boykin_practical_2005,boykin_brillouin-zone_2007,ku_unfolding_2010,popescu_effective_2010,popescu_extracting_2012,allen_recovering_2013,medeiros_effects_2014,medeiros_unfolding_2015,dirnberger_electronic_2021}, but requires specifying a primitive unit cell for comparison and forces aperiodic systems to be approximated as periodic \cite{nishi_band-unfolding_2017,matsushita_unfolding_2018,sanchez-ochoa_unfolding_2019,zhang_unfolded_2022}. The spectral function $A(\mathbf{k},E)= -\frac{1}{\pi}\textrm{Im}[\langle \mathbf{k} | G(E) |\mathbf{k}\rangle]$, defined through the system's Green's function projected into a plane wave basis $|\mathbf{k}\rangle$, provides a complementary description of band-like features and remains well-defined in aperiodic systems, but does not imply that the system exhibits a state, or approximate state, at $\mathbf{k}$ \cite{marsal_topological_2020,marsal_obstructed_2023,ciocys2024establishing,schirmann2024physical,rogalev2015fermi}.

Thus, despite their utility, both spectral weights and spectral functions are fundamentally indirect: they reveal band-like features, but neither guarantees the existence of states that are simultaneously localized in both energy and momentum. 
Mathematically, this limitation is rooted in the all-or-nothing nature of whether two operators share an eigenbasis. If a system's Hamiltonian $H$ commutes with some set of unitary translation operators $T_j$, $[H,T_j]=0$, its band structure is the joint spectrum of these operators $H |\psi_{n,\mathbf{k}} \rangle = E |\psi_{n,\mathbf{k}} \rangle$ and $T_j |\psi_{n,\mathbf{k}} \rangle = e^{i \mathbf{k} \cdot \mathbf{a}_j} |\psi_{n,\mathbf{k}} \rangle$, where $|\psi_{n,\mathbf{k}} \rangle$ is a shared eigenvector, i.e., the wavefunction of the $n$th band with wavevector $\mathbf{k}$ and $\mathbf{a}_j$ is the lattice vector associated with $T_j$. Instead, if $[H,T_j] \ne 0$, no such joint spectrum or  can be defined as the two operators do not share an eigenbasis.
Yet, recently, the mathematical field of operator algebras has developed the concept of multi-operator pseudospectra \cite{trefethen_pseudospectra_1997,trefethen_spectra_2005,jefferies2004NC_spectral_theory,LoringPseudospectra,Vasilescu2021Spectrum_Clifford_operators,loringLuWatson2021locality,Colombo2022Normal_operators_Clifford_modules,mumford2023numbers_and_beyond,Cerjan_Loring_Vides_2023,lin2024almost_commuting_and_measurement,berkolaiko_loring--schulz-baldes_2025}, which enables the definition of approximate joint eigenvectors between almost-commuting operators. In other words, if operators $A$ and $B$ satisfy $\| [A,B] \| \le \delta$ for some constant $\delta$ and where $\| \cdot \|$ is the $\ell^2$ operator norm, then approximate joint eigenvectors $|\phi\rangle$ can be found such that $A |\phi \rangle \approx \lambda_A |\phi \rangle$ and $B |\phi \rangle \approx \lambda_B |\phi \rangle$, and the possible degree of exact\-ness of these approximations is set by $\delta$. Thus, these techniques present the possibility of defining unfolded band-like structures that can smoothly interpolate between both periodic and aperiodic materials.

Here, we develop a direct generalization of dispersion relations using the quadratic pseudospectrum and derive bounds that quantify the extent to which the system exhibits an approximate joint eigenstate that is localized in energy and crystal momentum.
Moreover, we show that by incorporating both a system's unitary translation operators and its position operators into this multi-operator pseudospectrum, we can apply our framework to systems with open boundaries while suppressing edge effects, enabling its direct application to disordered, quasicrystalline, and other aperiodic systems without any other approximations. To demonstrate this band-unfolding approach's broad applicability, we show unfolded approximate band structures for both a Fibonacci chain \cite{Jagannathan_2021} and the 2D breathing honeycomb lattice \cite{noh_topological_2018}. Looking forward, our approach rooted in multi-operator pseudospectral methods enables the prediction of the momentum-space properties of materials regardless of the presence of sparse defects, incommensurate twist angles, magnetic fields, or other weak perturbations that break a system's periodicity.
 
\begin{figure}[t]
    \includegraphics[width=\linewidth]{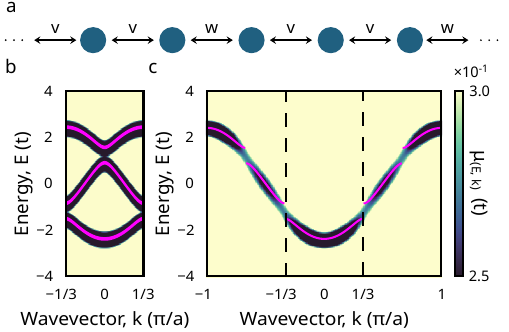}
    \caption{
        Band unfolding of the infinite trimerized lattice with a site-to-site separation of $a$, hopping amplitudes $ v = 0.8 t$ and $w = 1.2t$ where $t$ is the energy scale, and $\kappa = t$. 
        (a) Schematic of the trimerized lattice. (b) Exact folded band structure of this model. (c) Heatmap of the quadratic gap $\mu_{(E,k)}$ in units of $t$ with $k$ chosen in the range of the primitive Brillouin zone $k \in [-\pi,a,\pi/a)$. The analytic solution \cite{Zhang_Ren_Li_Ye_2021} for the unfolded band structure is superimposed in magenta. %For these parameters, $\| [H, T] \| = 0.57 t$.
    }
    \label{fig:ssh folded unfolded comparison}
\end{figure}

%%% Methods %%%
To illustrate our approach, we first consider an example with known band unfolding, a 1D monatomic chain with nearest neighbor couplings in which a perturbation has been added to every third coupling, see Fig.~\ref{fig:ssh folded unfolded comparison}a. Thus, if $T$ is the primitive translation operator that shifts each site to its neighbor on the right by the primitive lattice constant $a$, the perturbation means that $[H, T] \ne 0$. As translation by three sites leaves the system invariant, $[H, T^3] = 0$, the system possesses a folded band structure that consists of three bands in the supercell Brillouin zone $k \in (-\pi /3a, \pi/3a]$, see Fig.~\ref{fig:ssh folded unfolded comparison}b.

Nevertheless, for a weak perturbation such that the system's total Hamiltonian satisfies $\| [H, T] \| \lesssim t$, where $t$ is the Hamiltonian's energy scale, an approximate momentum-resolved structure in the primitive Brillouin zone $k \in (-\pi/a,\pi/a]$ can be found from the system's quadratic $\varepsilon$-pseudospectrum. First, a quadratic composite operator can be formed from $H$ and $T$ as
\begin{align} \label{eq:Q1d}
    \begin{split}
        Q_{(E,k)} \left(H,T\right) =  &\left( H - E \one \right)^2 \\
            + \kappa^2 &\left( T - e^{ika} \one \right)^\dagger \left( T - e^{ika} \one \right),
    \end{split}
\end{align}
where $\one$ is the identity. Here, $\kappa$ is a hyperparameter with units of energy that balances the spectral weights of $H$ and $T$ in $Q_{(E,k)}$, i.e., heuristically, it sets the relative tolerance between energy and momentum resolution. Note that by definition, $Q_{(E,k)}$ is both Hermitian and semi-positive. The quadratic $\varepsilon$-pseudospectrum is then defined in terms of the smallest eigenvalue of $Q_{(E,k)}$ as
\begin{equation}
    \Lambda_\varepsilon(H,T) = \{ (E,k) \; | \; \mu_{(E,k)}(H,T) \le \varepsilon  \},
\end{equation}
in which the quadratic gap
\begin{equation}
    \mu_{(E,k)}(H,T) = \sqrt{\min\left(\textrm{spec}[Q_{(E,k)}(H,T)]\right)}
\end{equation}
is the minimal eigenvalue's square root such that $\mu_{(E,k)}$ has units of energy and $\textrm{spec}(M)$ is the spectrum of $M$.

Intuitively, $Q_{(E,k)}$ is formed from the Hermitian squares of the eigenvalue equations associated with each constituent operator, in which $E$ and $k$ are inputs and act as ``eigenvalue guesses.'' Given such a set of eigenvalue guesses, $\mu_{(E,k)}$ then assesses whether the system exhibits an approximate plane-wave state with that $(E,k)$. More precisely, for $[H,T]\ne 0$, any normalized state in the system's Hilbert space $|\phi \rangle \in \mathcal{H}$, $\langle \phi | \phi \rangle = 1$ is guaranteed to deviate from simultaneously satisfying the constituent eigenvalue equations, with a minimum deviation set by the quadratic gap,
\begin{equation} \label{eq:estimate statement}
    \left( \| (H-E \one) |\phi \rangle \|^2 + \|(T-e^{ika} \one)|\phi\rangle \|^2 \right)^{\frac{1}{2}} \ge \mu_{(E,k)}(H,T).
\end{equation}
This minimum deviation is provably saturated by the quadratic composite operator's eigenstate associated with its smallest eigenvalue $Q_{(E,k)}|\phi^{(Q)} \rangle = \mu_{(E,k)}|\phi^{(Q)} \rangle$, see Supplemental Information \cite{SI}. Note, if $[H,T] = 0$, their joint spectrum directly yields the locations where $\mu_{(E,k)}=0$ through the left-hand side of Eq.~\eqref{eq:estimate statement}.
Thus, locations in $(E,k)$-space where $\mu_{(E,k)}$ is small indicate that the system supports an approximate state $|\phi^{(Q)} \rangle$ that satisfies $H|\phi^{(Q)}\rangle \approx E |\phi^{(Q)}\rangle$ and $T|\phi^{(Q)}\rangle \approx e^{ika}|\phi^{(Q)}\rangle$. Conversely, if $\mu_{(E,k)}$ is large, no such state approximately localized at $(E,k)$ exists.

In practice, the locus of $(E,k)$ that yield small $\mu_{(E,k)}$, i.e., the quadratic $\varepsilon$-pseudospectrum with small $\varepsilon$, can be viewed as an unfolded dispersion curve, identifying those locations in $(E,k)$-space where the system exhibits approximate plane wave states. In the case of the trimerized lattice, $\Lambda_\varepsilon(H,T)$ accurately reproduces both the locations of the bands as well as the band gaps introduced by the perturbation, see Fig.~\ref{fig:ssh folded unfolded comparison}c.

%%% Spatially Resolved %%%
A key advantage of using a multi-operator pseudospectral approach is that it enables a bulk-sensitive momentum-resolved description even in finite systems, without requiring periodic boundary conditions. In particular, it allows boundary-localized phenomena to be systematically suppressed, isolating the intrinsic, approximate bulk band structure. To do so, the quadratic composite operator can be expanded to contain additional operators that penalize the contributions of the system's boundaries,
\begin{align} \label{eq:QCO for spatially resolved}
    \begin{split} %To vertically center equation number
    Q_{(E,k,x)} ={} & \left( H - E \one \right)^2 \\
        & + \kappa_T^2 \left( T - e^{ika} \one \right)^\dagger \left( T - e^{ika} \one \right) \\
        & + \kappa_X^2 S \left( X - x \one \right)^2.
    \end{split}
\end{align}
Here, $X$ is the position operator, $\kappa_X$ is a hyperparameter with units of energy per length, and $S$ is a diagonal smoothing operator, e.g., so that $S( X - x \one)^2$ smoothly interpolates between $0$ near $x$ and $1$ in other locations. Thus, through the inclusion of $S( X - x \one)^2$, small values of the quadratic gap $\mu_{(E,k,x)}$ indicate the presence of approximate states localized at $(E,k)$ whose spatial support primarily resides within the domain centered on $x$ whose extent is chosen by the smoothing operator $S$. Note that for non-commuting operators, one is never guaranteed that locations with small $\mu_{(E,k,x)}$ exist, e.g., if the $S$-specified spatial domain has too restricted a volume to exhibit plane wave-like states.

\begin{figure}[t]
    \includegraphics[width=\linewidth]{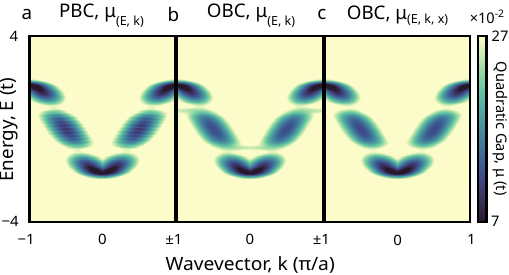}
    \caption{
        (a) Approximate unfolded band structure using $\mu_{(E,k)}$ of a finite trimerized lattice model with periodic boundary conditions (PBC). The finite lattice has $72$ unit cells of $3$ sites each, with $v = 0.8t$ and $w = 1.2t$, and $\kappa_T = 0.3t$. (b) Approximate unfolded band structure using $\mu_{(E,k)}$ of the same system, but with open boundary conditions (OBC).  (c) Approximate boundary-independent unfolded band structure using $\mu_{(E,k,x)}$ with $\kappa_X = 0.1t$ with OBC. For $x$ chosen in the lattice's center, the smoothing function is zero on sites $72$ through $144$, $1$ on sites $1$ through $36$ and $180$ through $216$, and smoothly interpolates between $0$ and $1$ in between, see Supplemental Information \cite{SI}.
    }
    \label{fig:ssh_resolved}
\end{figure}

To demonstrate this capability, we consider a finite trimerized lattice and compare the unfolded spectra obtained with and without the spatial resolution provided through $S( X - x \one)^2$.
Specifically, the approximate unfolded band structures of the system with periodic boundaries [Fig.~\ref{fig:ssh_resolved}a] and open boundaries [Fig.~\ref{fig:ssh_resolved}b] using Eq.~\eqref{eq:Q1d}, reveal that the trimerized lattice can exhibit boundary-localized states that manifest as partial flat bands in the unfolded band structure consistent with their spatial localization. Thus, this example shows that in finite systems, Eq.~\eqref{eq:Q1d} can exhibit features that stem from the system's boundaries. However, by including $S( X - x \one)^2$ in Eq.~\eqref{eq:QCO for spatially resolved} that penalizes the portions of the lattice away from the choice of $x$, the corresponding quadratic gap is mainly sensitive to the lattice's middle when $x$ is fixed in the middle and the boundary-localized state features are omitted in the approximate unfolded band structure regardless of lattice termination, see Fig.~\ref{fig:ssh_resolved}c.

%%% Fibonacci %%%
Having established that the quadratic pseudospectrum yields a boundary-independent, momentum-resolved description of finite systems, we now turn to a more stringent test: a system with no underlying periodicity whatsoever. To this end, we consider a one-dimensional quasiperiodic tight-binding model in which the coupling strengths alternate according to the Fibonacci sequence \cite{Jagannathan_2021}. Unlike the trimerized lattice, this system admits no nontrivial translation symmetry, as there is no nonzero translation under which the Hamiltonian is invariant. Consequently, Bloch’s theorem does not apply, and there is no well-defined crystal momentum nor a conventional band structure $E(k)$.

\begin{figure}[t]
    \centering
    \includegraphics[width=\linewidth]{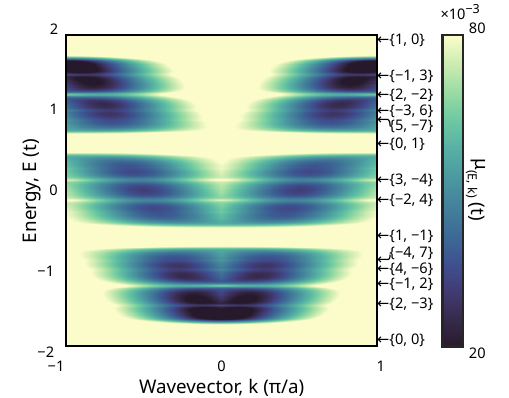}
    \caption{
        %Approximate dispersion for an open Fibonacci lattice calculated using the bulk-resolved quadratic gap $\mu_{(E,k,x)}$. The chain was formed using $12$ recursive substitutions yielding a total system length of $987$ sites \textcolor{red}{(tripled how? just gluing 3 copies together? Would you get essentially the same thing with a 13th substitution? Gluing 3 copies together is wrong, sorry if I suggested this.)}, in which the coupling coefficients are $v = 0.7t$ and $w = 1.0t$ with $t$ setting the energy scale. Pseudospectral calculations used $\kappa_T = 0.05 t$ and $\kappa_X = 0.1(t/a)$, where $a$ is the site-to-site spacing. Using $x$ in the system's center, the smoothing function selected for the sites $900$ through $2061$ \textcolor{red}{(replace if needed)}.
        %The predicted mini-gap locations labeled by $\{p,q\}$ are marked on the right-side of the approximate unfolded band structure. 
        Approximate dispersion for an open Fibonacci lattice calculated using the bulk-resolved quadratic gap $\mu_{(E,k,x)}$. The chain was formed using $12$ recursive substitutions yielding a total system length of $987$ sites, in which the coupling coefficients are $v = 0.7t$ and $w = 1.0t$ with $t$ setting the energy scale. Pseudospectral calculations used $\kappa_T = 0.05 t$ and $\kappa_X = 0.1(t/a)$, where $a$ is the site-to-site spacing. Using $x$ in the system's center and \( 14 \) substitutions to obtain a longer chain, the smoothing function selected for the sites $862$ through $1722$.
        The predicted mini-gap locations labeled by $\{p,q\}$ are marked on the right-side of the approximate unfolded band structure \cite{bellissard1992gap,Jagannathan_2021}. 
        %For the general story about such gap labeling schemes, see \cite{bellisard1992gapbook}. 
    }
    \label{fig:fibonacci}
\end{figure}

Here, we consider an open Fibonacci chain constructed recursively via the substitutions $w \mapsto v$ and $v \mapsto vw$ applied to the starting word $vvw$, with the resulting word specifying the sequence of coupling coefficients. As the number of substitution steps increases, the system approaches an infinite quasiperiodic limit characterized by a dense hierarchy of spectral gaps. Despite the absence of any exact translational symmetry, the quadratic pseusospectrum continues to identify approximate joint eigenstates of $H$ and $T$, thereby yielding an effective momentum-resolved description. To isolate bulk behavior, we incorporate the position operator as in Eq.~\eqref{eq:QCO for spatially resolved} and evaluate $\mu_{(E,k,x)}$ with $x$ chosen in the chain's interior. The resulting quadratic pseudospectrum reveals a structured dispersive feature interrupted by a hierarchy of gaps, see Fig.~\ref{fig:fibonacci}. These gaps align with the well-known gap-labeling scheme $\{q,p\}$ for the Fibonacci quasicrystal, where $q$ corresponds to a Chern number inherited from a higher-dimensional embedding \cite{bellissard1992gap,Jagannathan_2021}. Note that even though the open Fibonacci chain can exhibit in-gap edge states associated with the topological gap-labeling structure inherited from its higher-dimensional embedding, the corresponding features have been suppressed through the inclusion of $S(X - x\one)^2$ and choosing $x$ near the chain's center. In this way, the quadratic pseudospectrum provides a notion of band unfolding even in systems where no Bloch description exists, capturing in this case a dispersive envelope together with a structured hierarchy of spectral gaps.

%%% Breathing %%%
Approximate band unfolding using the quadratic pseudospectrum can be performed in higher dimensions as well, with an important caveat in the construction of $Q(H,\mathbf{T})$, where $\mathbf{T} = (T_1,...,T_m)$ are the included translation operators. In two or more dimensions, many space groups cannot be specified by a set of orthogonal lattice vectors with minimal lengths, e.g., in a 2D triangular lattice, there are three possible minimal-length lattice vectors that are equivalent under $C_3$, of which two can be chosen. When solving for a system's exact joint spectra with the corresponding translation operators, the information contained within the resulting band structure is independent of the choice of minimal-length lattice vectors. However, when using the quadratic pseudospectrum to find an approximate band structure, the omission of translation operator corresponding to a linearly dependent minimal-length lattice vector can have a deliterious effect. Consider the multi-dimensional generalization of Eq.~\eqref{eq:Q1d},
\begin{align} \label{eq:QCO for breathing}
    \begin{split} %To vertically center equati on number
    Q_{(E ,\mathbf{k} )} \left( H ,\mathbf{T}\right) = &\left( H - E \one \right)^2 \\
        + \kappa_T^2 \sum_{j = 1}^m &\left( T_j - e^{i\mathbf{k}\cdot \mathbf{a}_j} \one \right)^\dagger \left( T_j - e^{i\mathbf{k} \cdot \mathbf{a_j}} \one \right),
    \end{split}
\end{align}
where we are assuming that the same $\kappa_T$ is used for all of the translation operator terms for simplicity. If such a translation operator is omitted, the resulting eigenvalue deviation estimate, analogous to Eq.~\eqref{eq:estimate statement}, will only contain terms that measure the deviations due to the eigenvalue guess $\mathbf{k}$ along the chosen lattice vectors. Thus, any deviations that predominantly manifest along an unchosen minimal-length lattice vector will be de-emphasized, such that the resulting approximate unfolded band structure does not obey the correct symmetry. As such, to find an approximate unfolded band structure with the correct symmetry using the quadratic pseudospectrum, all rotationally related translation operators should be included.

\begin{figure}[t]
    \centering
    \includegraphics[width=\linewidth]{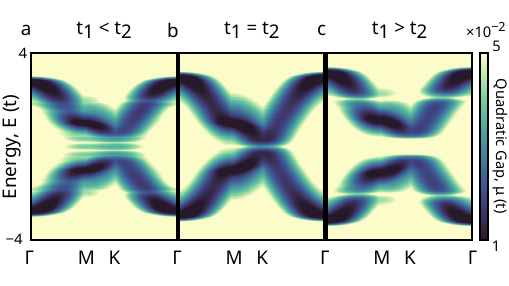}
    \caption{
        (a)-(c) Approximate unfolded band structures found using $\mu_{(E,\mathbf{k})}$ of the finite breathing honeycomb lattice (shown in the Supplement \cite{SI}), over a path in \( \mathbf{k} \)-space through high-symmetry points of the Brillouin zone. The finite system has $546$ total sites and the calculations use $\kappa_T = 0.1 t$. Three different realizations of the model are shown, with intra-hexagon couplings $t_1 = 0.8t$ and inter-hexagon couplings $t_2 = 1.2t$ (a), $t_1 = t_2 = t$ (b), and $t_1 = 1.2t$, $t_2 = 0.8t$ (c).
    }
    \label{fig:breathing vs graphene}
\end{figure}

To demonstrate the extension of this framework to higher dimensions, we consider a finite breathing honeycomb lattice, a minimal model originally considered for its ability to exhibit higher-order topological phases \cite{noh_topological_2018}. This model consists of a triangular lattice of hexagons, with potentially different coupling coefficients for the intra-hexagon ($t_1$) and inter-hexagon ($t_2$) connections. To ensure that the resulting approximate band structure possesses the correct rotational symmetry, all three primitive translation operators $T_1$, $T_2$, and $T_3 = T_2^\dag T_1$ are included in $Q_{(E ,\mathbf{k})}$. The resulting approximate dispersion surfaces show that the quadratic pseudospectrum reconstructs key features of the band structure, including Dirac points and gap openings, directly from finite systems without invoking Bloch's theorem, see Fig.~\ref{fig:breathing vs graphene}. Note that the open boundaries are discretizing the Dirac cones when $t_1 = t_2$ such that the finite system does not possess states with \(E= 0 \). Moreover, boundary-localized phenomena can also be found in the system's topological phase ($t_1<t_2$), manifesting as flat features in $\mu_{(E,\mathbf{k})}$, which can be confirmed by viewing the associated eigenstates $|\phi^{(Q)}\rangle$ of $Q_{(E ,\mathbf{k})}$, see Supplemental Information \cite{SI}. 

%%% Conclusion %%%
In conclusion, we have introduced a general framework for extracting momentum-resolved spectral structure using the quadratic pseudospectrum. By identifying the unfolded band structure with the locus of points in $(E,k)$-space where the quadratic gap is small, this approach provides a direct means of constructing band-like descriptions and associated approximate eigenstates even when exact translational symmetry is absent, such as in systems with disorder, defects, incommensurate modulations, and quasiperiodicity. Furthermore, by incorporating position operators into the composite framework, our approach can isolate bulk spectral features in finite systems while systematically suppressing boundary contributions. A distinguishing feature of this approach is its sensitivity to the degree of noncommutativity between operators, rather than only to qualitative distinctions such as commensurate versus incommensurate structure. In this sense, the quadratic pseudospectrum provides a continuous interpolation between periodic band theory and fully aperiodic systems, yielding a ``fuzzy'' momentum-space geometry that remains both numerically tractable and physically interpretable. 
%While both our pseudospectral approach and spectral functions can yield similar qualitative features when well-defined band-like structure is present, the pseudospectral formulation provides a direct criterion for the existence of approximate Bloch states and enables systematic isolation of bulk contributions.
Looking forward, this framework opens a pathway to analyzing momentum-resolved properties in systems such as incommensurate moir{\' e} materials, magnetic lattices with large or nonprimitive unit cells, quasicrystalline and aperiodic metamaterials, and chemically disordered alloys, where conventional unfolding approaches become cumbersome or require additional approximations.
\newline %So there is space before acknowledgements

\acknowledgements
We thank A.\ Grushin for helpful discussions regarding the Fibonacci chain.
C.B.\ acknowledges support from the Center for Integrated Nanotechnologies Summer Research Initiative, part of the Laboratory Directed Research and Development program at Sandia National Laboratories.
A.C.\ acknowledges support from the Laboratory Directed Research and Development program at Sandia National Laboratories.
This work was performed in part at the Center for Integrated Nanotechnologies, an Office of Science User Facility operated for the U.S.\ Department of Energy (DOE) Office of Science.
Sandia National Laboratories is a multimission laboratory managed and operated by National Technology \& Engineering Solutions of Sandia, LLC, a wholly owned subsidiary of Honeywell International, Inc., for the U.S. DOE's National Nuclear Security Administration under Contract No. DE-NA-0003525. 
R.K.\ acknowledges support from the Simons foudation. 
T.A.L.\ acknowledges support from the Army Research Office, Grant Number W911NF-25-1-0052. 
The views expressed in the article do not necessarily represent the views of the U.S. DOE or the United States Government.

\bibliography{refs}

@misc{SI,
	note={See Supplemental Information for a discussion of the derivation of Eq.~4 and related results, notes about the implementation of periodic boundary conditions for Fig.~1, the smoothing function used for selecting bulk phenomena, and boundary-localized states found using the quadratic composite operator for the breathing honeycomb lattice.}
}

@article{Zhang_Ren_Li_Ye_2021,
    title={Topological states in the super-SSH model},
    volume={29},
    rights={© 2021 Optica Publishing Group},
    ISSN={1094-4087},
    DOI={10.1364/OE.445301},
    abstractNote={The topological edge state distributes along the edge of a topological insulator which has advantages in prohibiting radiation and reflection in the evolution dynamics because of the topological protection property. The Su-Schrieffer-Heeger (SSH) model provides the simplest lattice configuration that supports topological edge states. Here, we investigate the properties of an extended SSH model &#x2013; super-SSH model &#x2013; with three sites in a unit cell for one-dimensional case and nine sites in a unit cell for two-dimensional case. Theoretical analysis and numerical simulation demonstrate that topological edge states and topological defect states are supported in the super-SSH model. This work extends the form of SSH model and may serve as a novel platform for developing photonic techniques based on topological phase transition.},
    number={26},
    journal={Optics Express},
    publisher={Optica Publishing Group},
    author={Zhang, Yiqi and Ren, Boquan and Li, Yongdong and Ye, Fangwei},
    year={2021},
    month=dec,
    pages={42827–42836},
}

@article{Jagannathan_2021,
    title={The Fibonacci quasicrystal: Case study of hidden dimensions and multifractality},
    volume={93},
    DOI={10.1103/RevModPhys.93.045001},
    author={Jagannathan, Anuradha},
    year={2021},
    month=nov,
    pages={045001}
}

@article{allen_recovering_2013,
	title = {Recovering hidden {Bloch} character: {Unfolding} electrons, phonons, and slabs},
	volume = {87},
	shorttitle = {Recovering hidden {Bloch} character},
	url = {http://link.aps.org/doi/10.1103/PhysRevB.87.085322},
	doi = {10.1103/PhysRevB.87.085322},
	number = {8},
	urldate = {2015-11-18},
	journal = {Phys. Rev. B},
	author = {Allen, P. B. and Berlijn, T. and Casavant, D. A. and Soler, J. M.},
	month = feb,
	year = {2013},
	pages = {085322},
}

@article{medeiros_effects_2014,
	title = {Effects of extrinsic and intrinsic perturbations on the electronic structure of graphene: {Retaining} an effective primitive cell band structure by band unfolding},
	volume = {89},
	shorttitle = {Effects of extrinsic and intrinsic perturbations on the electronic structure of graphene},
	url = {http://link.aps.org/doi/10.1103/PhysRevB.89.041407},
	doi = {10.1103/PhysRevB.89.041407},
	number = {4},
	urldate = {2015-11-18},
	journal = {Phys. Rev. B},
	author = {Medeiros, Paulo V. C. and Stafstr{\" o}m, Sven and Bj{\" o}rk, Jonas},
	month = jan,
	year = {2014},
	pages = {041407},
}

@article{Cerjan_Loring_Vides_2023,
    title={Quadratic pseudospectrum for identifying localized states},
    volume={64},
    ISSN={0022-2488, 1089-7658},
    DOI={10.1063/5.0098336},
    note={arXiv:2204.10450 [cond-mat, physics:math-ph, physics:physics, physics:quant-ph]},
    number={2},
    journal={Journal of Mathematical Physics},
    author={Cerjan, Alexander and Loring, Terry A. and Vides, Fredy},
    year={2023},
    month=feb,
    pages={023501}
}

@article{ashcroftMermin,
  title={Solid state physics},
  author={Ashcroft, Neil W and Mermin, N David},
  journal={Physics (New York: Holt, Rinehart and Winston) Appendix C},
  volume={1},
  year={1976}
}

@article{ku_unfolding_2010,
	title = {Unfolding {First}-{Principles} {Band} {Structures}},
	volume = {104},
	issn = {0031-9007, 1079-7114},
	url = {http://link.aps.org/doi/10.1103/PhysRevLett.104.216401},
	doi = {10.1103/PhysRevLett.104.216401},
	number = {21},
	urldate = {2016-10-31},
	journal = {Physical Review Letters},
	author = {Ku, Wei and Berlijn, Tom and Lee, Chi-Cheng},
	month = may,
	year = {2010},
}

@article{RevModPhys.90.015001,
  title = {Weyl and Dirac semimetals in three-dimensional solids},
  author = {Armitage, N. P. and Mele, E. J. and Vishwanath, Ashvin},
  journal = {Rev. Mod. Phys.},
  volume = {90},
  issue = {1},
  pages = {015001},
  numpages = {57},
  year = {2018},
  month = {Jan},
  publisher = {American Physical Society},
  doi = {10.1103/RevModPhys.90.015001},
  url = {https://link.aps.org/doi/10.1103/RevModPhys.90.015001}
}

@misc{Garcia_2025,
    title={Clifford and quadratic composite operators with applications to non-Hermitian physics},
    url={http://arxiv.org/abs/2410.03880},
    DOI={10.48550/arXiv.2410.03880},
    note={arXiv:2410.03880 [math-ph]},
    publisher={arXiv},
    author={Garcia, Jose J.},
    year={2025},
    month={September}
}

@incollection{lee2003smooth,
  title={Smooth manifolds},
  author={Lee, John M},
  booktitle={Introduction to smooth manifolds},
  pages={1--29},
  year={2003},
  publisher={Springer}
}

@article{ciocys2024establishing,
  title={Establishing coherent momentum-space electronic states in locally ordered materials},
  author={Ciocys, Samuel T and Marsal, Quentin and Corbae, Paul and Varjas, Daniel and Kennedy, Ellis and Scott, Mary and Hellman, Frances and Grushin, Adolfo G and Lanzara, Alessandra},
  journal={Nature communications},
  volume={15},
  number={1},
  pages={8141},
  year={2024},
  publisher={Nature Publishing Group UK London}
}

@article{schirmann2024physical,
  title={Physical properties of an aperiodic monotile with graphene-like features, chirality, and zero modes},
  author={Schirmann, Justin and Franca, Selma and Flicker, Felix and Grushin, Adolfo G},
  journal={Physical Review Letters},
  volume={132},
  number={8},
  pages={086402},
  year={2024},
  publisher={APS}
}

@article{rogalev2015fermi,
  title={Fermi states and anisotropy of Brillouin zone scattering in the decagonal Al--Ni--Co quasicrystal},
  author={Rogalev, VA and Gr{\"o}ning, O and Widmer, R and Dil, JH and Bisti, F and Lev, LL and Schmitt, T and Strocov, VN},
  journal={Nature communications},
  volume={6},
  number={1},
  pages={8607},
  year={2015},
  publisher={Nature Publishing Group UK London}
}

@article{bellissard1992gap,
  title={Gap labelling theorems for one dimensional discrete Schr{\"o}dinger operators},
  author={Bellissard, Jean and Bovier, Anton and Ghez, Jean-Michel},
  journal={Reviews in Mathematical Physics},
  volume={4},
  number={01},
  pages={1--37},
  year={1992},
  publisher={World Scientific}
}

@article{nuckolls_microscopic_2024,
	title = {A microscopic perspective on moiré materials},
	volume = {9},
	copyright = {2024 Springer Nature Limited},
	issn = {2058-8437},
	url = {https://www.nature.com/articles/s41578-024-00682-1},
	doi = {10.1038/s41578-024-00682-1},
	number = {7},
	urldate = {2025-12-11},
	journal = {Nat. Rev. Mater.},
	author = {Nuckolls, Kevin P. and Yazdani, Ali},
	month = jul,
	year = {2024},
	pages = {460--480},
}

@article{andrei_graphene_2020,
	title = {Graphene bilayers with a twist},
	volume = {19},
	copyright = {2020 Springer Nature Limited},
	issn = {1476-4660},
	url = {https://www.nature.com/articles/s41563-020-00840-0},
	doi = {10.1038/s41563-020-00840-0},
	number = {12},
	urldate = {2026-01-06},
	journal = {Nat. Mater.},
	author = {Andrei, Eva Y. and MacDonald, Allan H.},
	month = dec,
	year = {2020},
	pages = {1265--1275},
}

@article{he_moire_2021,
	title = {Moir{\' e} {Patterns} in {2D} {Materials}: {A} {Review}},
	volume = {15},
	issn = {1936-0851},
	shorttitle = {Moiré {Patterns} in {2D} {Materials}},
	url = {https://doi.org/10.1021/acsnano.0c10435},
	doi = {10.1021/acsnano.0c10435},
	number = {4},
	urldate = {2026-01-06},
	journal = {ACS Nano},
	author = {He, Feng and Zhou, Yongjian and Ye, Zefang and Cho, Sang-Hyeok and Jeong, Jihoon and Meng, Xianghai and Wang, Yaguo},
	month = apr,
	year = {2021},
	pages = {5944--5958},
}

@book{turek_electronic_2013,
	title = {Electronic {Structure} of {Disordered} {Alloys}, {Surfaces} and {Interfaces}},
	isbn = {978-1-4615-6255-9},
	publisher = {Springer Science \& Business Media},
	author = {Turek, Ilja and Drchal, V{\' a}clav and Kudrnovsk{\' y}, Josef and Sob, Mojm{\' i}r and Weinberger, Peter},
	month = nov,
	year = {2013},
}

@article{onida_electronic_2002,
	title = {Electronic excitations: density-functional versus many-body {Green}'s-function approaches},
	volume = {74},
	shorttitle = {Electronic excitations},
	url = {https://link.aps.org/doi/10.1103/RevModPhys.74.601},
	doi = {10.1103/RevModPhys.74.601},
	number = {2},
	urldate = {2026-01-06},
	journal = {Rev. Mod. Phys.},
	author = {Onida, Giovanni and Reining, Lucia and Rubio, Angel},
	month = jun,
	year = {2002},
	pages = {601--659},
}

@article{boykin_practical_2005,
	title = {Practical application of zone-folding concepts in tight-binding calculations},
	volume = {71},
	issn = {1098-0121, 1550-235X},
	url = {http://link.aps.org/doi/10.1103/PhysRevB.71.115215},
	doi = {10.1103/PhysRevB.71.115215},
	number = {11},
	urldate = {2016-10-31},
	journal = {Physical Review B},
	author = {Boykin, Timothy B. and Klimeck, Gerhard},
	month = mar,
	year = {2005},
}

@article{dirnberger_electronic_2021,
	title = {Electronic {State} {Unfolding} for {Plane} {Waves}: {Energy} {Bands}, {Fermi} {Surfaces}, and {Spectral} {Functions}},
	volume = {125},
	issn = {1932-7447},
	shorttitle = {Electronic {State} {Unfolding} for {Plane} {Waves}},
	url = {https://doi.org/10.1021/acs.jpcc.1c02318},
	doi = {10.1021/acs.jpcc.1c02318},
	number = {23},
	urldate = {2026-01-05},
	journal = {J. Phys. Chem. C},
	author = {Dirnberger, David and Kresse, Georg and Franchini, Cesare and Reticcioli, Michele},
	month = jun,
	year = {2021},
	pages = {12921--12928},
}

@article{boykin_brillouin-zone_2007,
	title = {Brillouin-zone unfolding of perfect supercells having nonequivalent primitive cells illustrated with a SiGe tight-binding parameterization},
	volume = {76},
	url = {https://link.aps.org/doi/10.1103/PhysRevB.76.035310},
	doi = {10.1103/PhysRevB.76.035310},
	number = {3},
	urldate = {2026-01-05},
	journal = {Phys. Rev. B},
	author = {Boykin, Timothy B. and Kharche, Neerav and Klimeck, Gerhard},
	month = jul,
	year = {2007},
	pages = {035310},
}

@article{popescu_extracting_2012,
	title = {Extracting \${E}\$ versus \${\textbackslash}stackrel\{{P}{\textbackslash}vec\}\{k\}\$ effective band structure from supercell calculations on alloys and impurities},
	volume = {85},
	url = {https://link.aps.org/doi/10.1103/PhysRevB.85.085201},
	doi = {10.1103/PhysRevB.85.085201},
	number = {8},
	urldate = {2026-01-05},
	journal = {Phys. Rev. B},
	author = {Popescu, Voicu and Zunger, Alex},
	month = feb,
	year = {2012},
	pages = {085201},
}

@article{popescu_effective_2010,
	title = {Effective {Band} {Structure} of {Random} {Alloys}},
	volume = {104},
	url = {https://link.aps.org/doi/10.1103/PhysRevLett.104.236403},
	doi = {10.1103/PhysRevLett.104.236403},
	number = {23},
	urldate = {2026-01-05},
	journal = {Phys. Rev. Lett.},
	author = {Popescu, Voicu and Zunger, Alex},
	month = jun,
	year = {2010},
	pages = {236403},
}

@article{medeiros_unfolding_2015,
	title = {Unfolding spinor wave functions and expectation values of general operators: {Introducing} the unfolding-density operator},
	volume = {91},
	shorttitle = {Unfolding spinor wave functions and expectation values of general operators},
	url = {https://link.aps.org/doi/10.1103/PhysRevB.91.041116},
	doi = {10.1103/PhysRevB.91.041116},
	number = {4},
	urldate = {2026-01-05},
	journal = {Phys. Rev. B},
	author = {Medeiros, Paulo V. C. and Tsirkin, Stepan S. and Stafstr{\" o}m, Sven and Bj{\" o}rk, Jonas},
	month = jan,
	year = {2015},
	pages = {041116},
}

@article{nishi_band-unfolding_2017,
	title = {Band-unfolding approach to moir{\textbackslash}'e-induced band-gap opening and {Fermi} level velocity reduction in twisted bilayer graphene},
	volume = {95},
	url = {https://link.aps.org/doi/10.1103/PhysRevB.95.085420},
	doi = {10.1103/PhysRevB.95.085420},
	number = {8},
	urldate = {2026-01-05},
	journal = {Phys. Rev. B},
	author = {Nishi, Hirofumi and Matsushita, Yu-ichiro and Oshiyama, Atsushi},
	month = feb,
	year = {2017},
	pages = {085420},
}

@article{sanchez-ochoa_unfolding_2019,
	title = {Unfolding method for periodic twisted systems with commensurate {Moiré} patterns},
	volume = {32},
	issn = {0953-8984},
	url = {https://doi.org/10.1088/1361-648X/ab44f0},
	doi = {10.1088/1361-648X/ab44f0},
	number = {2},
	urldate = {2026-01-05},
	journal = {J. Phys.: Condens. Matter},
	author = {S{\' a}nchez-Ochoa, F and Hidalgo, Francisco and Pruneda, Miguel and Noguez, Cecilia},
	month = oct,
	year = {2019},
	pages = {025501},
}

@article{zhang_unfolded_2022,
	title = {Unfolded band structures of photonic quasicrystals and moir{\textbackslash}'e superlattices},
	volume = {105},
	url = {https://link.aps.org/doi/10.1103/PhysRevB.105.165304},
	doi = {10.1103/PhysRevB.105.165304},
	number = {16},
	urldate = {2026-01-05},
	journal = {Phys. Rev. B},
	author = {Zhang, Yanbin and Che, Zhiyuan and Liu, Wenzhe and Wang, Jiajun and Zhao, Maoxiong and Guan, Fang and Liu, Xiaohan and Shi, Lei and Zi, Jian},
	month = apr,
	year = {2022},
	pages = {165304},
}

@article{matsushita_unfolding_2018,
	title = {Unfolding energy spectra of double-periodicity two-dimensional systems: {Twisted} bilayer graphene and \$\{{\textbackslash}mathrm\{{MoS}\}\}\_\{2\}\$ on graphene},
	volume = {2},
	shorttitle = {Unfolding energy spectra of double-periodicity two-dimensional systems},
	url = {https://link.aps.org/doi/10.1103/PhysRevMaterials.2.010801},
	doi = {10.1103/PhysRevMaterials.2.010801},
	number = {1},
	urldate = {2026-01-05},
	journal = {Phys. Rev. Mater.},
	author = {Matsushita, Yu-ichiro and Nishi, Hirofumi and Iwata, Jun-ichi and Kosugi, Taichi and Oshiyama, Atsushi},
	month = jan,
	year = {2018},
	pages = {010801},
}

@article{dubois_properties-_2012,
	title = {Properties- and applications of quasicrystals and complex metallic alloys},
	volume = {41},
	issn = {1460-4744},
	url = {https://pubs.rsc.org/en/content/articlelanding/2012/cs/c2cs35110b},
	doi = {10.1039/C2CS35110B},
	number = {20},
	urldate = {2026-01-07},
	journal = {Chem. Soc. Rev.},
	author = {Dubois, Jean-Marie},
	month = sep,
	year = {2012},
	pages = {6760--6777},
}

@article{steurer_twenty_2004,
	title = {Twenty years of structure research on quasicrystals. {Part} {I}. {Pentagonal}, octagonal, decagonal and dodecagonal quasicrystals},
	volume = {219},
	copyright = {De Gruyter expressly reserves the right to use all content for commercial text and data mining within the meaning of Section 44b of the German Copyright Act.},
	issn = {2196-7105},
	url = {https://www.degruyterbrill.com/document/doi/10.1524/zkri.219.7.391.35643/html?lang=en},
	doi = {10.1524/zkri.219.7.391.35643},
	number = {7},
	urldate = {2026-01-07},
	journal = {Zeitschrift für Kristallographie - Crystalline Materials},
	author = {Steurer, Walter},
	month = jul,
	year = {2004},
	pages = {391--446},
}

@article{tsai_icosahedral_2008,
	title = {Icosahedral clusters, icosaheral order and stability of quasicrystals—a view of metallurgy*},
	volume = {9},
	issn = {1468-6996},
	url = {https://doi.org/10.1088/1468-6996/9/1/013008},
	doi = {10.1088/1468-6996/9/1/013008},
	number = {1},
	urldate = {2026-01-07},
	journal = {Sci. Technol. Adv. Mater.},
	author = {Tsai, An Pang},
	month = apr,
	year = {2008},
	pages = {013008},
}

@book{jefferies2004NC_spectral_theory,
  title={Spectral Properties of Noncommuting Operators Lecture Notes in Mathematics 1843},
  author={Jefferies, Brian},
  year={2004},
  publisher={Springer-Verlag Berlin}
}

@article{LoringPseudospectra,
    AUTHOR = {Loring, Terry A.},
     TITLE = {{$K$}-theory and pseudospectra for topological insulators},
   JOURNAL = {Ann. Physics},
  FJOURNAL = {Annals of Physics},
    VOLUME = {356},
      YEAR = {2015},
     PAGES = {383--416},
      ISSN = {0003-4916},
   MRCLASS = {46L80 (46L85 65F99)},
  MRNUMBER = {3350651},
       DOI = {10.1016/j.aop.2015.02.031},
       URL = {http://dx.doi.org/10.1016/j.aop.2015.02.031},
}

@article{Vasilescu2021Spectrum_Clifford_operators,
    AUTHOR = {Vasilescu, Florian-Horia},
     TITLE = {Spectrum and analytic functional calculus for {C}lifford
              operators via stem functions},
   JOURNAL = {Concr. Oper.},
  FJOURNAL = {Concrete Operators},
    VOLUME = {8},
      YEAR = {2021},
    NUMBER = {1},
     PAGES = {90--113},
      ISSN = {2299-3282},
   MRCLASS = {47A60 (30A05 30G35 47A10)},
  MRNUMBER = {4281295},
       DOI = {10.1515/conop-2020-0115},
       URL = {https://doi.org/10.1515/conop-2020-0115},
}

@article{loringLuWatson2021locality,
  title={Locality of the windowed local density of states},
  author={Loring, Terry A and Lu, Jianfeng and Watson, Alexander B},
  journal={Numer. Math.},
  volume={156},
  number={2},
  pages={741--775},
  year={2024},
}

@book{mumford2023numbers_and_beyond,
  title={Numbers and the world: essays on math and beyond},
  author={Mumford, David},
  year={2023},
  publisher={American Mathematical Society}
}

@article{lin2024almost_commuting_and_measurement,
  title={Almost commuting self-adjoint operators and measurements},
  author={Lin, Huaxin},
  journal={arXiv preprint arXiv:2401.04018},
  year={2024}
}

@article{Colombo2022Normal_operators_Clifford_modules,
    AUTHOR = {Colombo, Fabrizio and Kimsey, David P.},
     TITLE = {The spectral theorem for normal operators on a {C}lifford
              module},
   JOURNAL = {Anal. Math. Phys.},
  FJOURNAL = {Analysis and Mathematical Physics},
    VOLUME = {12},
      YEAR = {2022},
    NUMBER = {1},
     PAGES = {Paper No. 25, 92},
      ISSN = {1664-2368,1664-235X},
   MRCLASS = {47B15 (47A10 47S05)},
  MRNUMBER = {4356506},
       DOI = {10.1007/s13324-021-00628-8},
       URL = {https://doi.org/10.1007/s13324-021-00628-8},
}

@misc{berkolaiko_loring--schulz-baldes_2025,
	title = {The {Loring}--{Schulz}-{Baldes} {Spectral} {Localizer} {Revisited}},
	url = {http://arxiv.org/abs/2512.21843},
	doi = {10.48550/arXiv.2512.21843},
	urldate = {2026-01-01},
	publisher = {arXiv},
	author = {Berkolaiko, Gregory and Shapiro, Jacob and White, Beyer Chase},
	month = dec,
	year = {2025},
	note = {arXiv:2512.21843 [math-ph]},
}

@article{noh_topological_2018,
	title = {Topological protection of photonic mid-gap defect modes},
	volume = {12},
	copyright = {2018 The Author(s)},
	issn = {1749-4893},
	url = {https://www.nature.com/articles/s41566-018-0179-3},
	doi = {10.1038/s41566-018-0179-3},
	urldate = {2018-06-22},
	journal = {Nat. Photonics},
	author = {Noh, Jiho and Benalcazar, Wladimir A. and Huang, Sheng and Collins, Matthew J. and Chen, Kevin P. and Hughes, Taylor L. and Rechtsman, Mikael C.},
	year = {2018},
	pages = {408},
}

@article{trefethen_pseudospectra_1997,
	title = {Pseudospectra of {Linear} {Operators}},
	volume = {39},
	issn = {0036-1445},
	url = {https://epubs.siam.org/doi/abs/10.1137/S0036144595295284},
	doi = {10.1137/S0036144595295284},
	number = {3},
	urldate = {2022-05-02},
	journal = {SIAM Rev.},
	author = {Trefethen, Lloyd N.},
	month = jan,
	year = {1997},
	pages = {383--406},
}

@book{trefethen_spectra_2005,
	title = {Spectra and {Pseudospectra}},
	isbn = {978-0-691-11946-5},
	url = {https://press.princeton.edu/books/hardcover/9780691119465/spectra-and-pseudospectra},
	urldate = {2022-05-02},
	publisher = {Princeton University Press},
	author = {Trefethen, Lloyd N. and Embree, M.},
	month = aug,
	year = {2005},
}

@article{marsal_topological_2020,
	title = {Topological {Weaire}–{Thorpe} models of amorphous matter},
	volume = {117},
	url = {https://www.pnas.org/doi/full/10.1073/pnas.2007384117},
	doi = {10.1073/pnas.2007384117},
	number = {48},
	urldate = {2026-04-30},
	journal = {Proceedings of the National Academy of Sciences},
	publisher = {Proceedings of the National Academy of Sciences},
	author = {Marsal, Quentin and Varjas, D{\' a}niel and Grushin, Adolfo G.},
	month = dec,
	year = {2020},
	pages = {30260--30265},
}

@article{marsal_obstructed_2023,
	title = {Obstructed insulators and flat bands in topological phase-change materials},
	volume = {107},
	url = {https://link.aps.org/doi/10.1103/PhysRevB.107.045119},
	doi = {10.1103/PhysRevB.107.045119},
	number = {4},
	urldate = {2026-04-30},
	journal = {Phys. Rev. B},
	author = {Marsal, Quentin and Varjas, Daniel and Grushin, Adolfo G.},
	month = jan,
	year = {2023},
	pages = {045119},
}

\end{document}

\title{Supplemental Material for Band Unfolding via the Quadratic Pseudospectrum}

\author{Christopher A.\ Bairnsfather}
\affiliation{Department of Mathematics, Purdue University, West Lafayette, Indiana, 47907, USA}

\author{Ralph M.\ Kaufmann}
\affiliation{Department of Mathematics, Purdue University, West Lafayette, Indiana, 47907, USA}
\affiliation{Department of Physics and Astronomy, Purdue University, West Lafayette, Indiana, 47907, USA}

\author{Terry A.\ Loring}
%\email[]{loring@math.unm.edu}
\affiliation{Department of Mathematics and Statistics, University of New Mexico, Albuquerque, New Mexico 87131, USA}

\author{Alexander Cerjan}
\email[]{awcerja@sandia.gov}
\affiliation{Center for Integrated Nanotechnologies, Sandia National Laboratories, Albuquerque, New Mexico 87185, USA}

\maketitle

\section{Comparison of the Technique Using the Momentum Space Resolved Hamiltonian vs. the Real Space Resolved Hamiltonian}

\begin{figure}[htp] 
    \centering
    \includegraphics[width=0.7\textwidth]{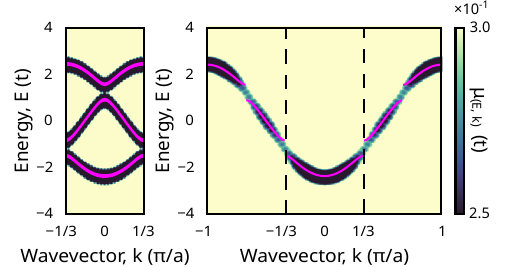}
    \caption{Band unfolding of the periodic trimerized lattice model with 24 unit cells of \( m = 3 \) sites each, a site-to-site separation of \( a \), hopping amplitudes \( v = 0.8t ,\, w = 1.2t \), and \( \kappa = t \). Folded bands of the original system (left) are shown for comparison against the approximate unfolded bands (right).
    }
    \label{fig:ssh with beading}
\end{figure}

The most straightforward way to obtain the approximate unfolded bands bands for the trimerized lattice model is to pick some large finite chain length, enforce periodic boundary conditions, and then apply the technique to the large finite matrix operators \( H \) and \( T \) describing this finite system. 
If this is done for a system with \( N \) total sites, the wavevector \( e^{ik} \) values must be complex \( N \)th roots of unity for consistency, i.e. we must have \( k = \frac{2\pi m}{N} \) for \( m = 0 ,\, 1 ,\, \dots ,\, N - 1 \). 
This approach leads to visual ``beads'' seen in the approximate bands found using the quadratic composite operator, as shown in Fig. \ref{fig:ssh with beading}. 
This is not necessarily a problem, since it reflects a real feature of the approximate joint eigenstates of the finite trimerized system with periodic boundary conditions, but it is not reflective of the \( N \to \infty \) limit. 
Of course, as \( N \to \infty \) the \( N \)th roots of unity become dense in the unit circle, and the visual beading disappears. To faithfully reproduce this limit, we can instead use the supercell Bloch Hamiltonian and Bloch-periodic translation operator
\begin{equation}
    H(k_{\textrm{sc}}) = \begin{bmatrix} 0 & v & w e^{-i k_{\textrm{sc}}} \\ v & 0 & v \\ w e^{i k_{\textrm{sc}}} & v & 0 \end{bmatrix}  
    \qquad
    T(k_{\textrm{sc}}) = \begin{bmatrix} 0 & 1 & 0 \\ 0 & 0 & 1 \\ e^{i k_{\textrm{sc}}} & 0 & 0 \end{bmatrix}
\end{equation}
where \( v \) and \( w \) are the intra- and inter-cell coupling constants.% and \( k_\text{ext} \) is the purported eigenvalue of \( T_\text{ext} \). 
This results in a 3D \( \left( E ,\, k ,\, k_{\textrm{sc}} \right) \) search space of eigenvalue guesses. 
The results shown in Fig.~1 of the main text minimize the quadratic gap values over the possible values of \( k_{\textrm{sc}} \) and display the resulting 2D heatmap.

\section{Error Bound and Plot of the Momentum Resolved Uncertainty}
Here we describe in more detail the bound on variance mentioned in the main paper. 
We need to introduce some notation for the expectation and variance of a not-necessarily Hermitian operator \( X \) in a state \( \ket{\psi} \), 
\begin{align} \label{eq:defn of expectation and variance for peanut proposition}
    \begin{split}
        E_{\ket{\psi}} \left[ X \right] &= \braket{\psi | X | \psi}, \\
        \Delta_{\ket{\psi}}^2 \left[ X \right] &= E_{\ket{\psi}} \left[ X^\dagger X \right] - \overline{E_{\ket{\psi}} \left[ X \right] }  E_{\ket{\psi}} \left[ X \right]. 
    \end{split}
\end{align}
The terms ``expectation'' and ``variance'' are suggestive, but there is no probability distribution of eigenvalues of \( X \) that gives them this meaning when \( X \) is non-Hermitian.
However, the only mathematical difference between this and the standard definition of the expectation and variance of an operator in a state is that, when we can assume that \( X \) is Hermitian, we can write \( X^2 \) in place of \( X^\dagger X \) and similarly for \( \left( E_{\ket{\psi}} \left[ X \right] \right)^2 \). 

We shall use \( \one \) to indicate the identity operator on \( \C^n \). 

\begin{prop} \label{prop:chris's corollary}
    Let \( H \) be a Hermitian operator and \( U_1 ,\, \dots ,\, U_d \) be unitary operators, all acting on the Hilbert space \( \C^n \). 
    Let \( E ,\, k_j \in \R \) for \( j = 1 ,\, \dots ,\, d \) and think of \( \left( E ,\, e^{ik_1} ,\, \dots ,\, e^{ik_d} \right) \) as a tuple of purported eigenvalues for the operators \( H ,\, U_1 ,\, \dots ,\, U_d \), respectively. 
    Let \( \kappa_1 ,\, \dots ,\, \kappa_d \geq 0 \). 
    Then the following are equal
    \begin{enumerate}[label=\roman*)] 
        \item \( \underset{\braket{\psi | \psi} = 1}{\min} \sqrt{ \braket{\psi | \left( H - E \one\right)^2 | \psi} + \displaystyle\sum_{j = 1}^d \kappa_j^2 \braket{\psi | \left( U_j - e^{i k_j} \one \right)^\dagger \left( U_j - e^{i k_j} \one \right) | \psi}} \),
        \item \( \underset{\braket{\psi | \psi} = 1}{\min} \sqrt{ \Delta_{\ket{\psi}}^2 \left[ H \right] + \left( E_{\ket{\psi}} \left[ H \right] - E \right)^2 + \displaystyle\sum_{j = 1}^d \kappa_j^2 \left( \Delta_{\ket{\psi}}^2 \left[ U_j \right] + \left\vert E_{\ket{\psi}} \left[ U_j \right] - e^{ik_j} \right\vert^2 \right) } \),
        \item The smallest singular value of \( M = \begin{bmatrix} H - E \one \\ \kappa_1 \left( U_1 - e^{ik_1} \one \right) \\ \vdots \\ \kappa_d \left( U_d - e^{ik_d} \one \right) \end{bmatrix} \), 
        \item The square root of the smallest eigenvalue of \( Q_{H ,\, U_1 ,\, \dots ,\, U_d} \left( E ,\, e^{i k_1} ,\, \dots ,\, e^{i k_d} \right) \). 
    \end{enumerate}
\end{prop}
\begin{proof}
    We adapt the proof found in \cite{Cerjan_Loring_Vides_2023} to the case where some of the operators are unitary; the argument there extends to this case with some adjustments. 

    For any operator \( X \) in the normalized state \( \ket{\psi} \) and any \( \lambda \in \C \) we compute 
    \begin{align} \label{eq:copied from CLV 2023 paper pretty much verbatim}
        \begin{split}
            \braket{\psi | \left( X - \lambda \one \right)^\dagger \left( X - \lambda \one \right) | \psi} &= \braket{ \psi | X^\dagger X | \psi} - \lambda \braket{ \psi | X^\dagger | \psi } - \overline{\lambda} \braket{ \psi | X | \psi } + \overline{\lambda} \lambda \\
            &= E_{\ket{\psi}} \left[ X^\dagger X \right] - \left( E_{\ket{\psi}} \left[ X \right] \right)^\dagger \left( E_{\ket{\psi}} \left[ X \right] \right) + \left( E_{\ket{\psi}} \left[ X \right] \right)^\dagger \left( E_{\ket{\psi}} \left[ X \right] \right) \\
            &- \lambda E_{\ket{\psi}} \left[ X^\dagger \right] + \overline{\lambda} E_{\ket{\psi}} \left[ X \right] + \left\vert \lambda \right\vert^2 \\
            &= \Delta_{\ket{\psi}}^2 \left[ X \right] + \left( E_{\ket{\psi}} \left[ X \right] - \lambda \right)^\dagger \left( E_{\ket{\psi}} \left[ X \right] - \lambda \right)
        \end{split}
    \end{align}
    where in the last line we use that \( \overline{ E_{\ket{\psi}} \left[ X \right] } = E_{\ket{\psi}} \left[ X^\dagger \right] \). 
    When \( X = H \) is Hermitian and \( \lambda = E \) is real, this gives us that 
    \begin{equation} \label{eq:hermitian case}
        \braket{\psi | \left( H - E \one \right)^2 | \psi} = \Delta_{\ket{\psi}}^2 \left[ H \right] + \left( E_{\ket{\psi}} \left[ H \right] - E \right)^2 
    \end{equation}
    and when \( X = U_j \) is unitary and \( \lambda = e^{i k_j} \) is unit complex, we instead obtain
    \begin{equation} \label{eq:unitary case}
        \braket{\psi | \left( U_j - e^{i k_j} \one \right)^\dagger \left( U_j - e^{i k_j} \one \right) | \psi} = \Delta_{\ket{\psi}}^2 \left[ U_j \right] + \left( E_{\ket{\psi}} \left[ U_j \right] - e^{i k_j} \right)^\dagger  \left( E_{\ket{\psi}} \left[ U_j \right] - e^{i k_j} \right). 
    \end{equation}
    We can then go term-by-term through i) to obtain ii), showing that they are equal. 

    For any matrix \( M \) the singular values of \( M \) are the square roots of the (necessarily non-negative real) eigenvalues of \( M^\dagger M \). 
    We note that  \( Q_{H ,\, U_1 ,\, \dots ,\, U_d} \left( E ,\, e^{i k_1} ,\, \dots ,\, e^{i k_d} \right) = M^\dagger M \). 
    Thus iii) = iv). 

    Finally, we recall that \( \sigma_\text{min} \left( M^\dagger M \right) = \underset{\braket{\psi | \psi} = 1}{\min} \braket{\psi | M^\dagger M | \psi} \) where \( \sigma_\text{min} \) denotes the smallest eigenvalue of a matrix. 
    We see from the definition of the quadratic composite operator that \( \braket{\psi | Q_{H ,\, U_1 ,\, \dots ,\, U_d} \left( E ,\, e^{i k_1} ,\, \dots ,\, e^{i k_d} \right) | \psi} \) is precisely i). 
    Thus all four quantities are equal. 
\end{proof}

\begin{rmk}
    Note that the proof does not really require the various \( U_j \) to be unitary, so even more generality is possible. 
    The extension from \cite{Cerjan_Loring_Vides_2023} requires modification of the notation, but once that is done, one may as well let the various \( U_j \) be arbitrary normal operators. 
    For results in this direction, especially as to the extent to which \( E_{\ket{\psi}} \left[ X \right] \) and \( \Delta_{\ket{\psi}} \left[ X \right] \) can be interpreted and used as true expectation values and variances in various cases, including the case of differing numbers of Hermitian vs. non-Hermition operators, see \cite{Garcia_2025}. 
\end{rmk}

\begin{figure}[htp]
    \centering
    \includegraphics[width=0.7\textwidth]{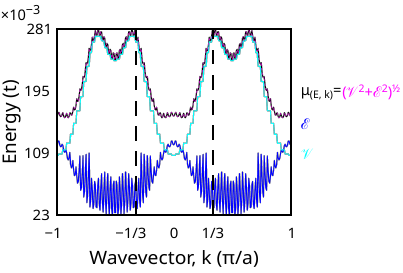}
    \caption{The partition of \( \mu_{(E,k)} \) (in black) into pieces due to deviation from expected value (in blue) and variance (in cyan) along a path \( E \left( k \right) \) is shown. 
    The sum of the expected value deviation contribution \( \mathcal{E} \) and the variance contribute \( \mathcal{V} \), in the sense of the inequality in (\ref{eq:inequality}), is given in magenta, and perfectly follows \( \mu_{(E,k)} \). 
    The path \( E \left( k \right) \) follows the dispersion curve for the \( v = w = 1 \) model where \( \left[ H ,\, T \right] = 0 \), but this is unnecessary; the equality would be perfect for any path. 
    All data is from the beaded version of the trimerized system considered above.} 
    \label{fig:Peanut Plot}
\end{figure}

Shown in Fig. \ref{fig:Peanut Plot} is a graph of the momentum-resolved uncertainty. 
We show the portion of \( \mu_{(E,k)}(H , T) \), with \( H \) and \( T \) corresponding to the beaded trimerized lattice model considered above, which is due to variance, denoted by \( \mathscr{V} \), and the portion due to deviation from expected value, denoted by \( \mathscr{E} \), and the total, for all points on the theoretical unfolded curve \( E \left( k \right) \) for a path that circumnavigates the unit cell. 
This illustrates that, not only is the inequality
\begin{align}
\begin{split} \label{eq:inequality}
    \mathscr{V} &= \underset{\braket{\psi | \psi} = 1}{\min} \sqrt{\Delta_{\ket{\psi}}^2 \left[ H \right] + \kappa_T^2 \Delta^2_{\ket{\psi}} \left[ T \right]} \\
    \mathscr{E} &= \underset{\braket{\psi | \psi} = 1}{\min} \sqrt{\left(E_{\ket{\psi}} \left[ H \right] - E \right)^2 + \kappa_T^2 \left| E_{\ket{\psi}} \left[ T \right] - e^{ik} \right|^2 } \\
    \mu^Q_{H ,\, T} \left( E ,\, k \right) &\leq \sqrt{\mathscr{V}^2 + \mathscr{E}^2}
\end{split}    
\end{align}
satisfied, but it is always saturated. 

\section{The Smoothing Function}
The smoothing function used for the spatially resolved band structure calculation is standard, as found in, e.g. \cite[Ch. 2]{lee2003smooth}. 
Let \( a < b < c < d \in \R \) and let \( f : \R \to \R \) be the mapping 
\begin{equation}
    f(t) = \begin{cases} 0 \,& t \leq 0 \\ e^{-\frac 1t} \,& t > 0 \end{cases}, 
\end{equation}
and use this to define \( g : \R \to \R \) as 
\begin{equation}
    g(t) = \dfrac{f(t)}{f(t) + f(1 - t)}. 
\end{equation}
Then 
\begin{equation}
    h(t) = 1 - g \left( \dfrac{t - a}{b - a} \right) g \left( \dfrac{d - t}{d - c} \right)
\end{equation}
is a smooth function which is identically \( 1 \) outside of \( \left(a ,\, d\right) \), is identically zero on \( \left[b ,\, c \right] \). 

In the main document, we choose \( a = 36, b = 72, c = 144 \), and \( d = 180 \) for our chain of total length \( 216 \).  
The resulting function \( f \) satisfies \( f(t) \equiv 1 \) on \( \left[ 1 ,\, 36 \right] \cup \left[ 180 ,\, 216 \right] \) and \( f(t) \equiv 0 \) on \( \left[ 72 ,\, 144 \right] \). 
The ``original chain'' is the length \( 72 \) segment on which \( f(t) \equiv 0 \). 
The diagonal operator \( S \) is defined via \( S_{jj} = f(j) \) for \( j = 1 ,\, 2 ,\, \dots ,\, 216 \). 

In the proof above of (\ref{eq:inequality}) we did not actually require the unitarity assumption (see the remark following the proof). 
The same argument proceeds verbatim if one of the \( U_j - e^{i k_d} \) operators is replaced by \( \sqrt{S} \left( X - x \one \right) \), which is diagonal and hence normal.
So the result applies to the spatially resolved case as well. 
As a result, we obtain a statement analogous to that of Proposition \ref{prop:chris's corollary} except with quantities of the form 
\begin{equation} \label{eq:chris's corollary with spatial resolution}
        \underset{\braket{\psi | \psi} = 1}{\min} \sqrt{ \braket{\psi | \left( H - E \one\right)^2 | \psi} + \displaystyle\sum_{j = 1}^{d - 1} \kappa_j^2 \braket{\psi | \left( U_j - e^{i k_j} \one \right)^\dagger \left( U_j - e^{i k_j} \one \right) | \psi} + \kappa_{d}^2 \braket{\psi | \left(  \sqrt{S} \left( X - x  \one \right) \right)^2 | \psi}}
\end{equation}
where \( x \) is the purported eigenvalue of the diagonal position operator \( X \).

\section{Boundary-localized States in Breathing Graphene}
\begin{figure}[htp] 
        \centering
        \includegraphics[keepaspectratio, width=0.7\textwidth]{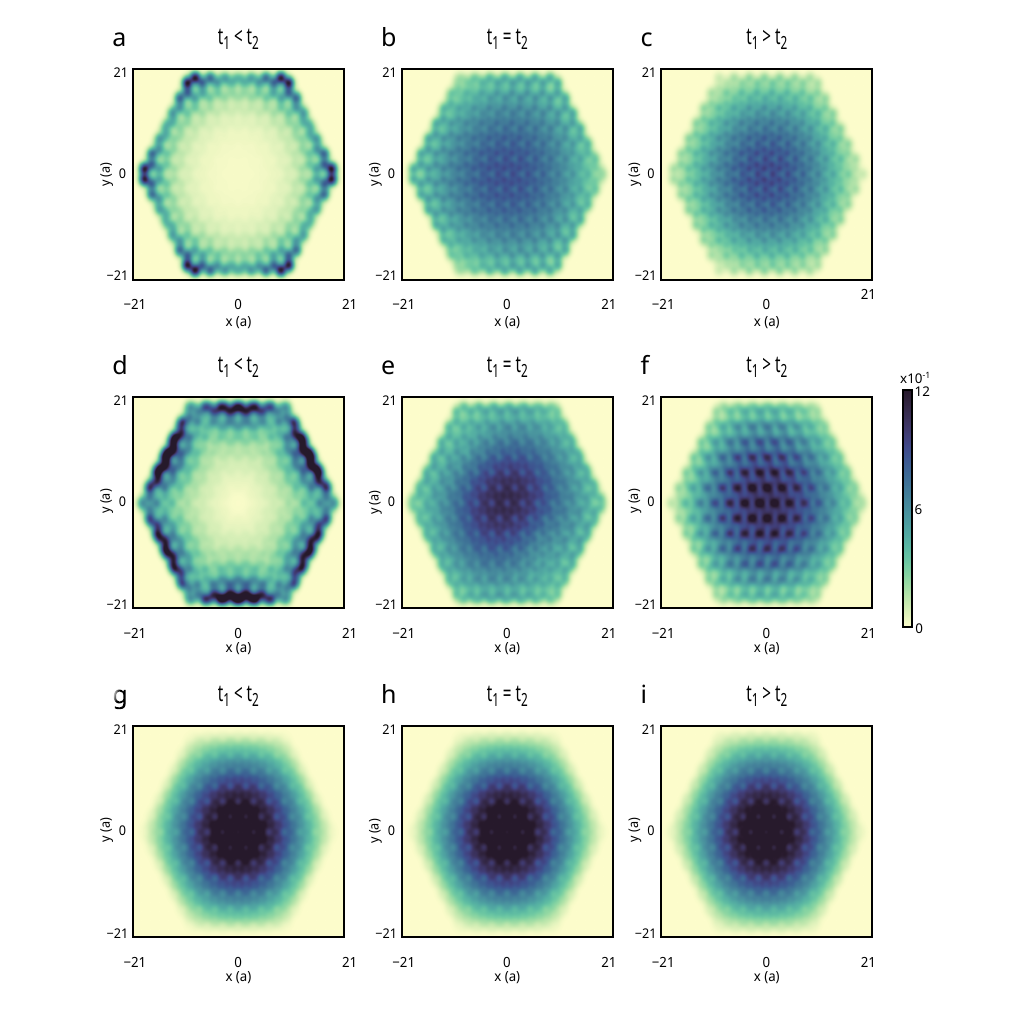}
        \caption{Shown are the best approximate joint eigenstates of \( H \) and the \( T_i \), for \( i = 1, 2, 3 \), at the location of a Dirac point. The top row is for \( E = 0 \), the middle for \( E = 0.15t \), and the bottom row is for \( E = 4t \). The systems in question are: \( t_1 < t_2 \) for a), d), and g); \( t_1 = t_2 \) (graphene) for b), e), and h); and \( t_1 > t_2 \) for c), f), and i). The color reflects the amplitude of the state at each site.}
        \label{fig:corner states}
\end{figure}

%Images of states in graphene, breathing, inverse breathing--rough form. These have not been Inkscaped. 
Shown in Fig. \ref{fig:corner states} are plots of some specific best approximate joint eigenstates for the various breathing graphene systems considered in the main document. 
The methodology is as follows--the best approximate joint eigenstate is found as a normalized eigenvector of \( Q_{H ,\, T_i} \left( E ,\, \mathbf{k} \right) = \left( H - E \one \right)^2 + \kappa_T^2 \sum_{i = 1}^3 \left( T_i - e^{i\mathbf{k}\cdot \mathbf{a}_i} \one \right)^\dagger \left( T_i - e^{i\mathbf{k} \cdot \mathbf{a_i}} \one \right) \) corresponding to the smallest eigenvalue.
Its amplitude at each site is extracted and used as the strength of a small Gaussian centered at each site. 
The sum of these Gaussians is plotted to visualize the distribution of the state in real space. 
We have chosen \( E = 0 \), \( E = 0.15t \), and \( E = 4t \) and \( \mathbf{k} \) the location of one of the Dirac points of graphene. 
In the breathing case, we find a well-localized corner state at \( E = 0 \) as one would expect since this system is a higher-order topological insulator \cite{noh_topological_2018}.
We can also find an edge state at \( E = 0.15t \). 
In the other cases, such topological states are not found.
In the case of graphene (\( t_1 = t_2 \)) at \( E = 0 \) we see a state strongly delocalized in real space, i.e. a state strongly localized in \( \mathbf{k} \)-space at the chosen Dirac point. 
In every case, we do not see a localized state in either momentum space or real space at \( E = 4t \) (far above the Dirac point).

\bibliography{refs}